\documentclass[prb,superscriptaddress,twocolumn,preprintnumbers,amsmath,amssymb]{revtex4}

\def\SRO{Sr$_2$RuO$_4$}
\def\RO{RuO$_2$}

\usepackage{graphicx}
\usepackage{dcolumn}
\usepackage{bm}

\begin{document}

\preprint{Journal of the Physical Society of Japan {\bf 73}, 1313 (2004)}

\title{Determination of the Superconducting Gap Structure in All Bands\\ of the Spin-Triplet Superconductor Sr$_2$RuO$_4$}

\author{K.~Deguchi}
\email{deguchi@scphys.kyoto-u.ac.jp}
\affiliation{Department of Physics, Graduate School of Science, Kyoto University, Kyoto 606-8502, Japan}

\author{Z.Q.~Mao}
 \altaffiliation{Present address: Physics Department, Tulane University, 2001 Percival Stern, New Orleans, LA 70118, USA.}
\affiliation{Department of Physics, Graduate School of Science, Kyoto University, Kyoto 606-8502, Japan}

\author{Y.~Maeno}
\affiliation{Department of Physics, Graduate School of Science, Kyoto University, Kyoto 606-8502, Japan}
\affiliation{Kyoto University International Innovation Center, Kyoto 606-8501, Japan}

\date{\today}

\begin{abstract}
We have obtained strong experimental evidence for the full determination of the superconducting gap structure in all three bands of the spin-triplet superconductor Sr$_2$RuO$_4$ for the first time. We have extended the measurements of the field-orientation dependent specific heat to include conical field rotations consisting of in-plane azimuthal angle $\phi$-sweeps at various polar angles $\theta$ performed down to 0.1 K. Clear 4-fold oscillations of the specific heat and a rapid suppression of it by changing $\theta$ are explained only by a compensation from two types of bands with anti-phase gap anisotropies with each other. The results indicate that the active band, responsible for the superconducting instability, is the $\gamma$-band with the lines of gap minima along the [100] directions, and the passive band is the $\alpha$- and $\beta$-bands with the lines of gap minima or zeros along the [110] directions in their induced superconducting gaps. We also demonstrated the scaling of the specific heat for the field in the {\it c}-direction, which supports the line-node-like gap structures running along the $k_z$ direction.
\end{abstract}

\maketitle

\section{Introduction}

The unconventional superconductivity, realized in high-$T_{\rm c}$ cuprates, organics, heavy-fermion intermetallic compounds, ruthenate, etc., has become one of the most actively studied topics of condensed matter physics today. Since the discovery of its superconductivity,~\cite{Maeno1} the layered ruthenate \SRO\ has generated extensive research activities.~\cite{Mackenzie1,Maeno2} What make \SRO\ special among many unconventional superconductors are the availabilities of detailed knowledge of its Fermi-liquid properties, as well as of ultra-high-purity single crystals with the residual resistivity less than 100 n$\Omega$cm.~\cite{Mao3} In particular, its Fermi surface is relatively simple and consists of three nearly-cylindrical sheets of $\alpha$-, $\beta$-, and $\gamma$-bands with specific {\it d}-electron orbital characters. Fermi-surface parameters of all three sheets, reflecting strong electron correlations, are experimentally quantified in detail.~\cite{Mackenzie3,Bergemann} Therefore, there is a prospect of the underlying physics of its unconventional superconductivity be understood in unprecedented detail and depth. Obviously, the determination of the symmetry and anisotropic structure of the order parameter, the superconducting gap, is a crucial step toward identification of its pairing mechanism.

	The superconductivity of \SRO\ has pronounced unconventional features such as: the invariance of the spin susceptibility across its superconducting (SC) transition temperature $T_{\rm c}$,~\cite{Ishida1,Duffy} appearance of spontaneous internal field,~\cite{Luke} evidence for two-component order parameter from field distribution in the vortex lattice,~\cite{Kealey} absence of a Hebel-Slichter peak,~\cite{Ishida3} and the strong dependence of $T_{\rm c}$ on nonmagnetic impurities.~\cite{Mackenzie2,Mao1} These features are coherently understood in terms of spin-triplet {\it p}-wave superconductivity with the vector order parameter being $\bm{d}(\bm{k})=\hat {\bm{z}}{\mathit \Delta}_{0}(k_x + {\rm i}k_y)$,~\cite{Rice} in contrast to the spin-singlet {\it d}-wave pairing known in high $T_{\rm c}$ cuprates. This order parameter represents the paired spin state with $S = 1$ and $S_z = 0$, and the orbital wave function with $L_z = +1$, called the chiral {\it p}-wave state. The above simple vector order parameter leads to the gap ${\mathit \Delta}(\bm{k})={\mathit \Delta}_{0}(k_x^2 + k_y^2)^{1/2}$, which is isotropic because of the quasi-two dimensionality of the Fermi surface. Exponential temperature dependence of quasiparticle (QP) excitations is expected in this case.

	However, a number of experimental results with very high quality crystals revealed the power-law temperature dependence of QP excitations. In particular, the specific heat~\cite{Nishizaki} $C/T \propto T$ , the thermal conductivity~\cite{Tanatar1} $\kappa/T \propto T$ and NQR relaxation rate~\cite{Ishida4} $1/T_1 \propto T^3$, in addition to the ultrasonic attenuation rate~\cite{Lupien} $\alpha$ and variation of the penetration depth~\cite{Bonalde} ${\mathit \Delta} \lambda$, all indicated the presence of the lines of nodes, of zeros, or of gap minima in the superconducting gap. Nevertheless, magnetothermal conductivity measurements for determination of the SC gap structure, with the applied field rotated within the \RO\ plane down to 0.35 K, have indicated little anisotropy within the planes.~\cite{Tanatar2,Izawa1} In spite of a number of detailed studies of the SC gap structure, the wave function of Sr$_2$RuO$_4$ are thus still controversial.

	The main outstanding issues on the unconventional superconductivity of Sr$_2$RuO$_4$ at present may be summarized as follows~\cite{Mackenzie1}: (1) full characterization of the SC gap structure (i.e. orbital state) in all bands, (2) direct observation of odd parity of the orbital state by a phase-sensitive method (3) clarification of the details of the spin state in the spin-triplet pairing state especially under magnetic fields, (4) confirmation of the ``chirality" due to time reversal symmetry breaking in the SC state and searches for novel phenomena associated with chirality, and (5) understanding of the multi-phase superconducting phases and the critical field limiting.~\cite{Deguchi1,Mao2,Yaguchi} Furthermore, the assignment of the dominant band in producing the superconductivity by taking the orbital dependent superconductivity (ODS) into account~\cite{Agterberg1}, as well as the examination of the role of the incommensurate AF fluctuation to pairing in the triplet channel, is crucial for the determination of the pairing mechanism.

	We report in this paper a new development to resolve the issue (1). Measurements of the spectra of QPs induced by magnetic fields serve as a powerful approach for this purpose. The field-orientation dependent specific heat is a direct measure of the QP density of states (DOS) and thus a powerful probe of the SC gap structure.~\cite{Vekhter,Won3,Won4,Miranovic,Park} Indeed, in our previous report, we revealed the emergence of in-plane anisotropy below about 0.3 K in the SC gap of Sr$_2$RuO$_4$ from the field-orientation dependence of the specific heat.~\cite{Deguchi2} We identified the multi-gap superconductivity with the active band $\gamma$, which has a modulated SC gap with a minimum along the [100] direction.

	In the present study, we extended the measurements of the field-orientation dependence of the specific heat to the field direction covering all polar and azimuthal angles. We also found the scaling of electronic specific heat in magnetic fields perpendicular to the \RO\ plane. As a result, we have determined the superconducting gap structure of the spin-triplet superconductor Sr$_2$RuO$_4$ in all bands for the first time. We will examine and select the pertinent theoretical models compatible with the present observation and discuss the effective pairing interaction as well as the roles of the incommensurate AF fluctuation in $\alpha$- and $\beta$-bands.

\section{Experiment}
Single crystals of \SRO\ were grown by a floating-zone method in an infrared image furnace.~\cite{Mao3} After specific-heat measurements on two crystals, the sample with $T_{\rm c} = 1.48$ K, close to the estimated value for impurity and defect free specimen ($T_{\rm c0}=1.50$ K),~\cite{Mackenzie2,Mao1} was chosen for detailed study. This crystal was cut and cleaved from the single crystalline rod, to a size of $2.8 \times 4.8 \;{\rm mm}^{2}$ in the $ab$-plane and $0.50 \;{\rm mm}$ along the $c$-axis. An x-ray rocking curve of the sample shows the characteristics of a single crystal of high quality; the diffraction peak width [full width at half maximum (FWHM)] was comparable to that of a Si crystal (with FWHM of $0.06 ^{\circ}$) in our diffractometer. The directions of the tetragonal crystallographic axes of the sample were determined by x-ray Laue pictures. The side of the crystal was intentionally misaligned from the [110] axis by $16^{\circ}$.

	The field-orientation dependence of the specific heat was measured by a relaxation method with a dilution refrigerator. The details of the measurement system, consisting of two orthogonally-arranged SC magnets and a mechanical rotating stage for the dewar, are described elsewhere.~\cite{Deguchi3} In the present study, we mainly operate the system with constant polar angle $\theta$ and varing azimuthal angle $\phi$ with a misalignment ${\mathit \Delta}\theta$ no greater than $0.01^{\circ}$. It is important to maintain a constant $\theta$ during this conical variation of the field direction, because the large $H_{\rm c2}$ anisotropy in $\theta$. The in-plane and inter-plane field alignment relative to the crystallographic axis was carried out, based on the known in-plane anisotropy of $H_{\rm c2}$ at low temperatures.~\cite{Deguchi1,Mao2,Yaguchi} For referring to the direction of the applied magnetic field with respect to the crystallographic axes, we shall introduce the polar angle $\theta$, for which $\theta = 0^{\circ}$ corresponds to the [001] direction, and the azimuthal angle $\phi$, for which $\phi = 0^{\circ}$ with $\theta = 90^{\circ}$ corresponds to the [100] direction. It is to be noted that the tetragonal symmetry of the crystal structure is known to conserve down to temperatures as low as 110 mK.~\cite{Gardner}\\

\section{Results and Discussion}

\subsection{Scaling of the specific heat for $\bm{H}\parallel [001]$ }

\begin{figure}[t]
    \begin{center}
\includegraphics[width=7.5cm,clip]{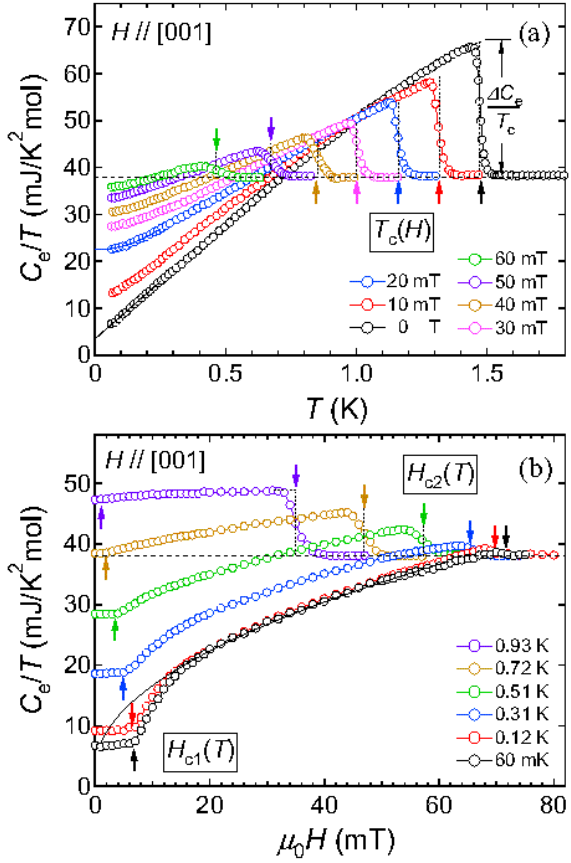}
    \end{center}
\caption{(a) Temperature dependence of the electronic specific heat divided by temperature $C_{\rm e}/T$ at several fields for $\bm{H} \parallel [001]$. The line at zero field is a fit of the low temperature $C_{\rm e}/T$ with $\gamma_{0} + \alpha T$. (b) Field dependence of $C_{\rm e}/T$ at several temperatures for $\bm{H} \parallel [001]$. The line is a fit with $\gamma_{0} + A\sqrt{H}$ for the data at 60 mK .}
\label{fig:THC}
\end{figure}

The objective of this section is to present the QP DOS in the vortex state of Sr$_2$RuO$_4$, when a magnetic field is applied perpendicular to the conducting plane, $\bm{H} \parallel [001]$. From such information, we try to determine the dimension of nodes or of gap minima: {\it lines} or {\it points}. Figure~\ref{fig:THC}(a) shows the temperature dependence of the electronic specific heat divided by temperature $C_{\rm e}/T$ at several fields. The electronic specific heat $C_{\rm e}$ in magnetic fields was obtained after subtraction of the phonon contribution with a Debye temperature of 410 K. The temperature dependence of $C_{\rm e}/T$ at low temperatures is proportional to $T$ down to 50 mK, and this temperature dependence suggests lines of nodes or of gap minima in the SC gap.~\cite{Nishizaki} Note that $C_{\rm e}/T$ at zero field linearly extrapolates to the residual electronic specific-heat coefficient $\gamma_0$ from about 0.5 K with $C_{\rm e}/T = \gamma_0 + \alpha T$: $\gamma_0 = 3.8$ mJ/K$^2$mol and $\alpha = 48.8$ mJ/K$^3$mol. The result of $\gamma_0/\gamma_{\rm N} = 0.1$ guarantees that this sample is clean enough, where $\gamma _{\rm N}$ is the electronic specific heat coefficient in the normal state: $\gamma_{\rm N} = 37.8$ mJ/K$^{2}$mol.
 
	To identify the dominant source of the DOS at low temperatures, we estimate the impurity scattering rate divided by the SC gap amplitude $\hbar{\mathit \Gamma}/{\mathit \Delta}_{0}=0.0098$ for $T_{\rm c}=1.48$ K from the Abrikosov-Gorkov equation $\ln (T_{\rm c0}/T_{\rm c})=\Psi (1/2+\hbar{\mathit \Gamma}/2\pi k_{\rm B}T_{\rm c})-\Psi (1/2)$. $\Psi$ is a digamma function, $T_{\rm c0}=1.50$ K is a maximum  $T_{\rm c}$ for the disorder-free material,~\cite{Mackenzie2} and we approximate $2{\mathit \Delta}_{0}/k_{\rm B}T_{\rm c0}=3.53$ (BCS). This result is in good agreement with the residual DOS estimated from the unitarity limit: $\gamma_0/\gamma_{\rm N} \simeq \sqrt{\hbar{\mathit \Gamma}/{\mathit \Delta}_{0}}$.~\cite{Balatsky,Suzuki} The condition of the superclean regime $\frac{\hbar{\mathit \Gamma}}{{\mathit \Delta}_{0}}\ll \frac{T }{T_{\rm c}}< \frac{H}{H_{\rm c20}}$ can be satisfied above $H_{\rm c1}$, where $H_{\rm c20} \equiv H_{\rm c2}(T=0\;{\rm K}) = 71$ mT for ${\bm H} \parallel c$ is estimated from $H_{\rm c2}(T)$ determined from the specific heat in Fig.~\ref{fig:THC}(a) and (b). Thus we conclude that the effect of impurity scattering is negligible. For $\frac{T }{T_{\rm c}}< \frac{H}{H_{\rm c20}}$, QPs are dominated by those induced by magnetic fields; in the mixed state, the QP energy spectrum $E_{\bm{k}}$ is affected and becomes $E^{\prime}_{\bm{k}} = E_{\bm{k}}-\delta \omega$ by the Doppler shift $\delta \omega = \hbar \bm{k} \cdot \bm{v}_s$, where $\bm{v}_s$ is the superfluid velocity around the vortices and $\hbar \bm{k}$ is the QP momentum.  In the case of $\delta \omega \geq {\mathit \Delta}(\bm{k})$, this energy shift gives rise to a finite DOS at the Fermi level. This leads to the field dependence of $C_{\rm e}/T \propto$ DOS being proportional to $\sqrt{H}$ at low temperature for the SC gap with lines of nodes or of gap minima (i.e. for the QP DOS of $\frac{N(E)}{N_{\rm F}} \propto \frac{E}{{\mathit \Delta}_0}$, where $E$ is the energy of a QP)~\cite{Volovik1}: $\frac{C_{\rm e}}{\gamma _{\rm N}T} \propto \frac{E_H}{{\mathit \Delta}_0}$ at zero temperature, where $E_{H}={\mathit \Delta}_0\sqrt{H/H_{\rm c20}}$ is the energy scale associated with the Doppler shift $\delta \omega$.~\cite{Vekhter} $N_{\rm F}$ is DOS at the Fermi level in the normal state.
\begin{figure}[t]
    \begin{center}
\includegraphics[width=7.5cm,clip]{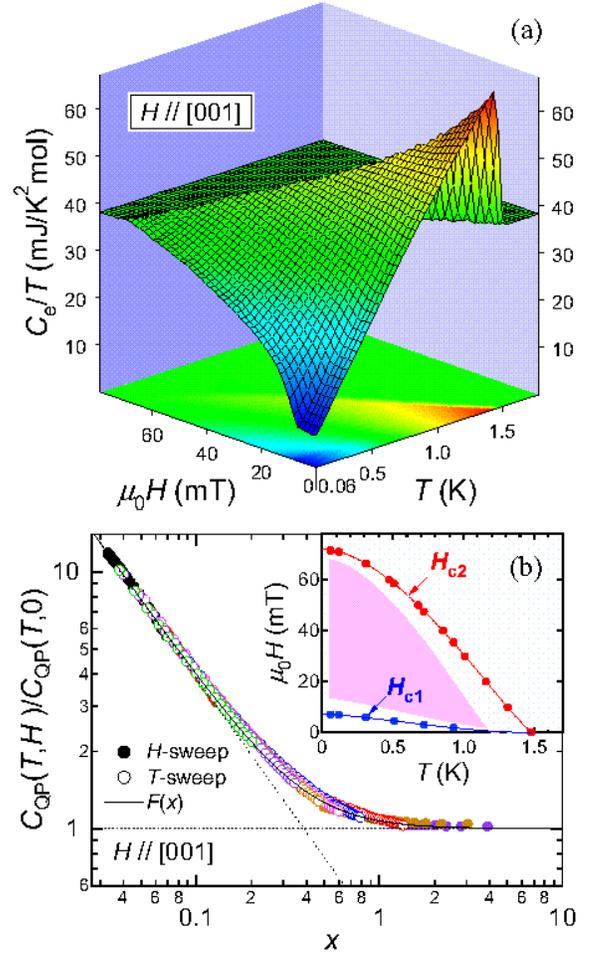}
    \end{center}
\caption{(a) Experimental data of $C_{\rm e}/T$ for $\bm{H} \parallel [001]$, as a function of field strength and temperature. A contour plot is shown on the bottom $H-T$ plane, with the same color scale as the 3D plot. (b) Parameter $x = a(T/T_c)/\sqrt{H/H_{c20}}$ dependence of the normalized QP contribution $C_{\rm QP}(T,H)/C_{\rm QP}(T,0)$ for $\bm{H} \parallel [001]$ with the same color as Fig.~\ref{fig:THC}(a) and (b), where $C_{\rm QP}(T,H)$ is defined as $C_{\rm e}(T,H) - \gamma_{0}T$. The inset shows $H$-$T$ phase diagram for $\bm{H} \parallel [001]$. The scaling of $C_{\rm QP}(T,H)/C_{\rm QP}(T,0)$ is carried out in the shaded region.}
\label{fig:SCAL}
\end{figure}

	Figure~\ref{fig:THC}(b) shows the field dependence of $C_{\rm e}/T$ at several temperatures for $\bm{H} \parallel [001]$, measured after zero-field cooling. At low temperatures, the specific heat behaves like $C_{\rm e}/T \propto \sqrt{H}$. The deviation from the  $\sqrt{H}$ law for $H < 14$ mT is most likely due to entrance into the Meissner state ($H_{\rm c1}(0)\sim 7$ mT). Well above $H_{\rm c1}$, field induced QPs become dominant and $C_{\rm e}/T $ has little temperature dependence below 0.1 K as shown in Fig.~\ref{fig:THC}(b). If the data at 60 mK is chosen as the limit of  $T/T_{\rm c} \rightarrow 0$, the field dependence of $C_{\rm e}/T$ is described as $\gamma_{0} + A\sqrt{H}$ with $A = 4.3$ mJ/K$^2$T$^{1/2}$mol. This field dependence also indicates nodes or of gap minima in the SC gap.

	Recently, it has been shown that the field induced QP DOS should yield general scaling relations of various thermodynamic and transport properties.~\cite{Simon,Won2} When the SC gap has lines of nodes or of gap minima in the field direction (i.e. lines of nodes or of gap minima $\parallel c$-axis and ${\bm H} \parallel c$-axis), the QP DOS is a simple function of $E/\sqrt{H}$. In this case, the scaling function has been explicitly calculated for the specific heat. The scaling relations are functions of a single reduced parameter $x = \frac{2}{v_{\rm F}}\sqrt{\frac{\phi_{0}}{\pi}}\frac{T}{\sqrt{H}} = a\frac{T}{T_{\rm c}}\sqrt{\frac{H_{\rm c20}}{H}}$, where $\phi_0$ is the flux quantum, $v_{\rm F}$ is the Fermi velocity in the conducting plane and $a$ is a constant of the order of unity.~\cite{Won2} Experimentally, this scaling of the specific heat has been demonstrated in high $T_{\rm c}$ cuprate superconductors.~\cite{Wang} To focus on the QP excitation only by the temperature and magnetic field, we define $C_{\rm QP}(T,H)\equiv C_{\rm e}(T,H) - \gamma_{0}T$. Then $C_{\rm QP}(T,H)/C_{\rm QP}(T,0)$ would be expressed by the scaling function $F(x)$ for the SC gap with lines of nodes or of gap minima,~\cite{Won2} where $F(x)$ is well approximated with the empirical interpolating function $F_{\rm int}(x)$.~\cite{Wang}
\begin{eqnarray}
F(x) = \frac{3}{2\pi^2}\frac{1}{x^3}\int_0^{\infty}ds s^2g(s){\rm sech}^2\left(\frac{s}{2x}\right),
\end{eqnarray}
\begin{eqnarray}
g(s) = \left\{
\begin{array}{lll}
 \frac\pi4 s(1+ \frac1{2 s^2}) & \mbox{$(s \ge 1)$}\\
\\
 \frac34 \sqrt{1-s^2}+ \frac{1}{4s}(1+2 s^2)\arcsin s & \mbox{$(s \le 1)$}
\end{array}
\right.,
\end{eqnarray}
\begin{eqnarray}
F_{\rm int}(x) = \sqrt{1 + \left(\frac{4\pi}{27\zeta (3)}\frac{1}{x}\right)^2}.
\end{eqnarray}
$F(x)$ and $F_{\rm int}(x)$ has two regimes, depending on whether the thermal energy $\frac{k_{\rm B}T}{{\mathit \Delta}_0} \simeq \frac{T}{T_{\rm c}}$ is small ($F(x),F_{\rm int}(x)\rightarrow \frac{4 \pi}{27\zeta(3)}\frac{1}{x}$ for $x \ll 1$, $C_{\rm QP} = AT\sqrt{H}$ for $T \rightarrow 0$) or large ($F(x),F_{\rm int}(x)\rightarrow 1$ for $x \gg 1$, $C_{\rm QP}=\alpha T^2$ in zero field) compared to the energy scale of the Doppler shift $\frac{\delta \omega}{{\mathit \Delta}_0} \simeq \frac{E_{H}}{{\mathit \Delta}_0} = \sqrt{\frac{H}{H_{\rm c20}}}$. The asymptotic form of $C_{\rm QP}(T,H)$ is given by
\begin{eqnarray}
C_{\rm QP}(T,H) = C_{\rm QP}(T,0)F(x) \simeq C_{\rm QP}(T,0)F_{\rm int}(x)\\
= \left\{
\begin{array}{lll}
a\frac{4 \pi \alpha T_{\rm c}}{27\zeta(3)}T \sqrt{\frac{H}{H_{\rm c20}}} & \mbox{$(x \ll 1,\; \frac{T}{T_{\rm c}}<\frac{1}{2})$} \\
\\
\alpha T^2 &\mbox{$(x \gg 1,\; \frac{T}{T_{\rm c}}<\frac{1}{2})$}
\end{array}
\right.,
\end{eqnarray}
where $C_{\rm QP}(T,0)$ approaches $\alpha T^2$ for $\frac{T}{T_{\rm c}}<\frac{1}{2}$ in Sr$_2$RuO$_4$.

	Figure~\ref{fig:SCAL}(a) is a 3-dimensional (3D) representation of $C_{\rm e}/T$ for the [001] field direction, as a function of field and temperature. The figure is constructed from data in Fig.~\ref{fig:THC}(a) and (b). Provided that the SC gap structure of \SRO has lines of nodes or of gap minima, but not points of those, $C_{\rm QP}(T,H)/C_{\rm QP}(T,0)$, deduced from this full map of the specific heat, would well be described by the scaling function $F(x)$ with the reduced parameter $x = aT/T_{\rm c}\sqrt{H_{\rm c20}/H}$ in the range of $0<T<T_{\rm c}$ and $H_{\rm c1}(T)<H<H_{\rm c2}(T)$.

	Figure~\ref{fig:SCAL}(b) shows the $x$ dependence of the normalized QP contribution $C_{\rm QP}(T,H)/C_{\rm QP}(T,0)$ of the electronic specific heat for $\bm{H} \parallel [001]$. Indeed the data obtained by $H$-sweep at constant $T$ and $T$-sweep at constant $H$ collapse onto a single curve in a wide range in the $H$-$T$ phase diagram, shown in the inset of Fig.~\ref{fig:SCAL}(b) as the shaded region, and the curve is well described by $F(x)$ even in the crossover region $x \approx 1$. The constant $a$ is determined from $a\frac{4 \pi \alpha T_{\rm c}}{27\zeta(3)} = A = 4.3$ mJ/K$^2$T$^{1/2}$mol in the limit of $x \ll 1$, where $\zeta(3)$ is 1.202..., and results in $a = 0.8$, indeed an order of unity. Such demonstration of the scaling strongly supports the existence of {\it lines} of nodes, of zeros, or of gap minima. In addition, the observation of $C_{\rm e}/T \propto \sqrt{H}$ up to near $H_{\rm c2}$ is also consistent with the chiral $p$-wave state.~\cite{Ichioka} Note that although it might be possible to explain these features by the single band model with lines of nodes or of gap minima, it is not so simple because the features of multi-band superconductivity emerges only by tilting the field direction to near $ab$-plane, as discussed in next section.

\subsection{Field-orientation dependence\\ of the specific heat}

In the previous section, we establish the scaling of $C_{\rm e}/T$ with the single parameter $x \propto T/\sqrt{H}$ for $\bm{H} \parallel [001]$, which indicates the existence of lines of nodes or of gap minima in the SC gap. In this section, we present the field-orientation dependent specific heat, from which we try to determine the direction of lines of nodes or of gap minima. Let us first present the results for $\theta = 90^{\circ}$~\cite{Deguchi2} with additional analysis. We will then present the new results for $\theta \neq 90^{\circ}$.
\begin{figure}[b]
    \begin{center}
\includegraphics[width=8.7cm,clip]{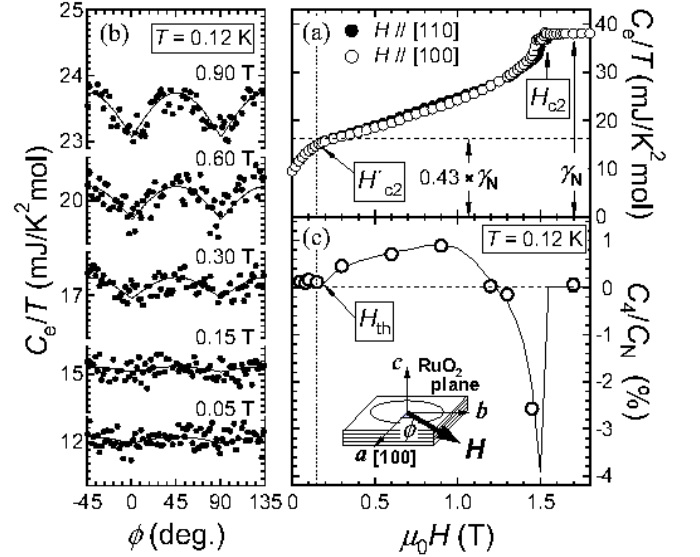}
    \end{center}
\caption{(a) Field dependence of $C_{\rm e}/T$ at $T = 0.12$ K for $\bm{H} \parallel [100]$ (open circle) and $\bm{H} \parallel [110]$ (closed circle). (b) The in-plane field-orientation dependence of $C_{\rm e}/T$ at several fields. The solid lines are fits with $f_{4}(\phi)$ given in the text. (c) Field dependence of the normalized 4-fold oscillation amplitude $C_{\rm 4}/C_{\rm N}$. The points are evaluated from the fitting to the oscillatory data in Fig.~\ref{fig:PRE}(b), while the lines from the difference in $C_{\rm e}$ between $\bm{H} \parallel [110]$ and $\bm{H} \parallel [100]$ in Fig.~\ref{fig:PRE}(a). Two methods yield consistent results.}
\label{fig:PRE}
\end{figure}

	Figure~\ref{fig:PRE}(a) shows the field dependence of $C_{\rm e}/T$ for ${\bm H} \parallel ab$-plane at $T = 0.12$ K. $C_{\rm e}/T$ increases sharply up to the characteristic field $\mu_{0}H^{\prime}_{\rm c2} = 0.15$ T and then increases more slowly for higher field. This behavior is more pronounced at 60 mK,~\cite{Deguchi2} and naturally explained by the presence of two kinds of the gap.~\cite{Nishizaki} In view of the QP excitations in ODS,~\cite{Agterberg1} both active and passive bands are effective below $H^{\prime}_{\rm c2}$ at low temperature region, but only the active band is important above $H^{\prime}_{\rm c2}$ and/or high temperatures since most of the QPs have already been excited for the passive bands. On the basis of the different orbital characters of the three Fermi surfaces ($\alpha, \beta,$ and $\gamma$),~\cite{Mackenzie3,Bergemann} the amplitudes of the gap functions ${\mathit \Delta}_{\alpha \beta }$ and ${\mathit \Delta}_{\gamma}$ are expected to be significantly different and the normalized DOS of those bands are $\frac{N_{\alpha \beta }}{N_{\rm F}} = 0.43$ and $\frac{N_{\gamma }}{N_{\rm F}} = 0.57$. The plateau on the field dependence of $C_{\rm e}/T$ at low temperature divides $\gamma_{\rm N}$ ($\propto N_{\rm F}$) reasonably into $0.43 \times \gamma_{\rm N}$ recovered at low field and $0.57 \times \gamma_{\rm N}$ recovered at high field. Consequently we conclude that the active band which retains strong superconductivity up to high field is the $\gamma$-band mainly derived from the in-plane $d_{xy}$ orbital of Ru $4d$ electrons.

	As mentioned in the previous section, the QP energy spectrum is affected by the Doppler shift in the mixed state for the superfluid velocity around the vortices, and consequently this energy shift gives rise to a finite DOS at the Fermi level where the local energy gap becomes smaller than the amplitude of the Doppler shift.~\cite{Volovik1} Since the superfluid velocity $\bm{v}_s$ is perpendicular to the magnetic field $\bm{H}$, the Doppler shift $\delta \omega$ becomes zero for $\bm{k} \parallel \bm{H}$. Thus the generation of QPs by magnetic fields at nodes or at gap minima is suppressed for $\bm{H} \parallel$ node or gap minima directions, and $C_{\rm e}/T$ takes minima for these directions.~\cite{Vekhter}

	Figure~\ref{fig:PRE}(b) shows the field-orientation dependence of $C_{\rm e}/T$ for ${\bm H} \parallel ab$-plane at $T = 0.12$ K. The absence of a 2-fold oscillatory component in the raw data at $\theta = 90^{\circ}$ guarantees that the in-plane field alignment is accurate during the azimuthal-angle $\phi$ rotation. For the field range $0.15$ T $<\mu_{0}H < 1.2$ T, where the QPs in the active band $\gamma$ are the dominant source of in-plane anisotropy of $C_{\rm e}(\phi)$, a {\it non-sinusoidal} 4-fold angular variation approximated as $C_{\rm e}(\phi) = C_{\rm 0} + C_{\rm 4}f_{\rm 4}(\phi)$ with $f_{\rm 4}(\phi) = 2|{\sin}2\phi| -1$ is observed. Since we observe (i) no angular variation above $H_{\rm c2}$ in the normal state and (ii) a 4-fold angular variation with the phase inverse to that due to the in-plane $H_{\rm c2}$ anisotropy below $H_{\rm c2}$ in the SC state, we conclude that the 4-fold oscillations at low temperature originate from the anisotropy in the SC gap.~\cite{Deguchi2} In addition, cusp-like features at the minima ($\phi = \frac{\pi}{2}n$, $n$: integer) are attributable to the strong reduction of QP excitations for the nodal direction parallel to $\bm{H}$ since the Doppler shift becomes $\delta \omega = 0$ for a purely two-dimensional gap structure.~\cite{Vekhter} On the other hand, strong $k_z$ dependence of the gap function would enhance the QP excitations even for the nodal direction parallel to $\bm{H}$, so that the cusp-like features would have been strongly suppressed. Therefore, in the active band $\gamma$, the gap structure with vertical line nodes or gap minima along the [100] directions are promising. In addition, from the consistency with observed value of ${\mathit \Delta}C_{\rm e}/\gamma_{\rm N}T_{\rm c}$, which mainly originates from the active band $\gamma$, we can deduce the order parameter $\bm{d}(\bm{k})=\hat {\bm{z}}{\mathit \Delta}_{0}({\sin}ak_x + {\rm i}{\sin}ak_y)$ with gap minima for the active band.~\cite{Deguchi2}

	Figure~\ref{fig:PRE}(c) shows the field dependence of $C_{\rm 4}/C_{\rm N}$ for ${\bm H} \parallel ab$-plane at $T = 0.12$ K. $C_{\rm 4}/C_{\rm N}$ is the normalized angular variation amplitude, where $C_{\rm N} = \gamma_{\rm N}T$ is the electronic specific heat in the normal state. At low fields ($\mu_{0}H \leq 0.15$ T) and low temperatures ($T \leq 0.3$ K) where QPs on both the active and passive bands are important, the oscillation anisotropy rapidly decreases. We define the threshold field $H_{\rm th}$ for the strong reduction of $C_{\rm 4}/C_{\rm N}$ in Fig.~\ref{fig:PRE}(c). While the previous experimental study~\cite{Deguchi2} has resolved the directions of gap minima in the active band $\gamma$, there still remains two types of possibilities predicted for the passive bands $\alpha$ and $\beta$: (A) horizontal line nodes~\cite{Zhitomirsky,Annett1,Annett2,Koikegami} and (B) vertical lines of nodes or of gap minima~\cite{Nomura2,Nomura3,Yanase1,Yanase2}.

	In view of the gap structure in the active band, this steep reduction of $C_{\rm 4}$ may be explained with the gap minima ${\mathit \Delta}_{\rm min}$ of  $\bm{d}(\bm{k})=\hat {\bm{z}}{\mathit \Delta}_{0}({\sin}ak_x + {\rm i}{\sin}ak_y)$: at low field (Doppler shift $\delta \omega \leq {\mathit \Delta}_{\rm min}$) the 4-fold oscillations cannot occur, whereas above the threshold field $H_{\rm th}$ ($\delta \omega \geq {\mathit \Delta}_{\rm min}$) the 4-fold oscillations will be observed. In this case, the gap structure of both (A) and (B) are still possible in passive bands, since these features may be explained with only the active band except for the power-law dependence of $C_{\rm e}/T \propto T$. Note that according to Fig.~\ref{fig:PRE} it is assumed in this scenario that $H_{\rm th}$ characterizing the gap minima of the active band happens to be nearly the same as $H^{\prime}_{\rm c2}$ characterizing the passive band.

	In view of ODS, the existence of vertical lines of nodes or of gap minima in the passive bands may give rise to 4-fold oscillations in $C_{\rm e}(\phi)$ which are out of phase with the 4-fold oscillations originating from the active band. In this case, the 4-fold oscillations would be suppressed below the characteristic field $H^{\prime}_{\rm c2}$, where both active and passive bands contribute to the $C_{\rm e}(\phi)$ oscillation. This idea requires strong in-plane anisotropy of the gap in passive bands and lead to the theoretical models (B).

	We have extended the specific heat measurements to cover all $\theta$ and $\phi$, in order to determine the gap structure in passive bands. Now, let us focus on the relation between the characteristic field $H^{\prime}_{\rm c2}$ of ODS and the threshold field $H_{\rm th}$ of steep $C_{\rm 4}$ reduction. First, we trace the change of $H^{\prime}_{\rm c2}$ by tilting the polar angle $\theta$. Next, we investigate the azimuthal angle $\phi$ dependence of $C_{\rm e}$ at several polar angle $\theta$, fixing the temperature and the {\it reduced} magnetic field $H/H_{c2}(\theta)$. In particular, we kept the reduced field such that $H_{\rm th}$ is always exceeded: $\frac{\delta \omega}{{\mathit \Delta}_0} \geq \frac{{\mathit \Delta}_{\rm min}}{{\mathit \Delta}_0} \simeq \frac{\delta \omega_{\rm th}}{{\mathit \Delta}_0}$ i.e. $\frac{H}{H_{\rm c2}(\theta)} \geq \frac{H_{\rm th}(\theta)}{H_{\rm c2}(\theta)} \equiv \frac{H_{\rm th}(90^{\circ})}{H_{\rm c2}(90^{\circ})}=\frac{0.15 \; {\rm T}}{1.5 \; {\rm T}}= 0.1$, where averaged $\delta \omega$ can be estimated from $\frac{\delta \omega}{{\mathit \Delta}_0} \sim \frac{E_{H}}{{\mathit \Delta}_0} = \sqrt{H/H_{\rm c20}(\theta)}$ and $\sqrt{H_{\rm th}(\theta)/H_{\rm c20}(\theta)} \sim \frac{\delta \omega_{\rm th}}{{\mathit \Delta}_0} \simeq \frac{{\mathit \Delta}_{\rm min}}{{\mathit \Delta}_0}$ = constant in the case that $H_{\rm th}$ is determined only by gap minima.~\cite{Vekhter,Won3,Won4} In the case that $C_{\rm 4}$ reduction is caused solely by gap minima in the active band, the polar angle $\theta$ dependence of $C_{\rm 4}$ is expected to be similar to that of single band models with vertical lines of nodes under the condition of $\delta \omega \geq {\mathit \Delta}_{\rm min}$. In that case $C_{\rm 4}(\theta)$ is not expected to be strongly affected by the evolution of the characteristic field $H^{\prime}_{\rm c2}$ of ODS. On the other hand, in the case that $C_{\rm 4}$ reduction is caused mainly by compensation of anti-phase gap anisotropies between active and passive bands, $C_{\rm 4}(\theta)$ is expected to be deeply related to the evolution of $H^{\prime}_{\rm c2}$. Especially, $C_{\rm 4}$ will rapidly be suppressed for $H \leq H^{\prime}_{\rm c2}$, for which anisotropic QP excitations can occur in both bands.
\begin{figure}[t]
    \begin{center}
\includegraphics[width=8cm,clip]{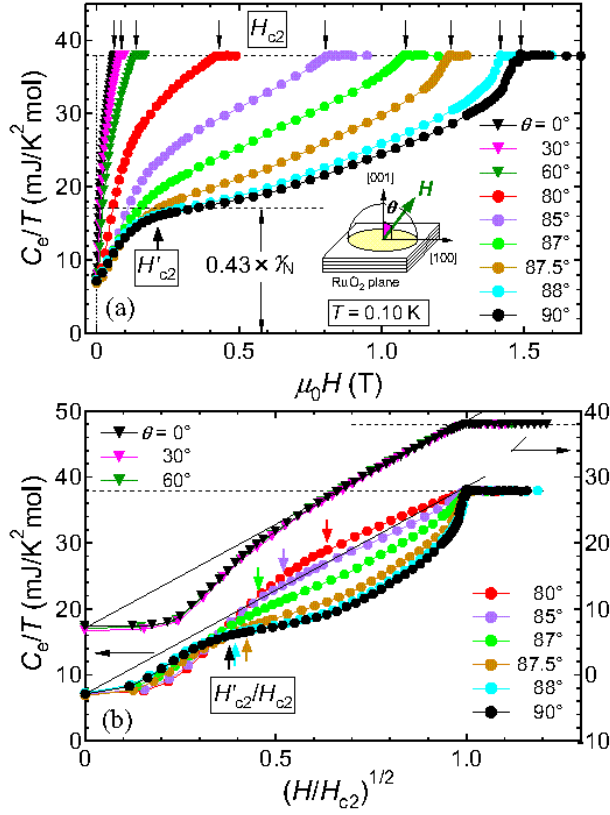}
    \end{center}
\caption{(a) Field dependence of  $C_{\rm e}/T$ at several polar angle $\theta$ from the [001] to [100] directions, fixing the temperature $T = 0.10$ K and the azimuthal angle $\phi = 0^{\circ}$. (b) $C_{\rm e}/T$ vs $\sqrt{H/H_{\rm c2}(\theta)}$at several polar angle $\theta$ from the [001] to [100] directions.}
\label{fig:THETA}
\end{figure}
\begin{figure}[t]
    \begin{center}
\includegraphics[width=8cm,clip]{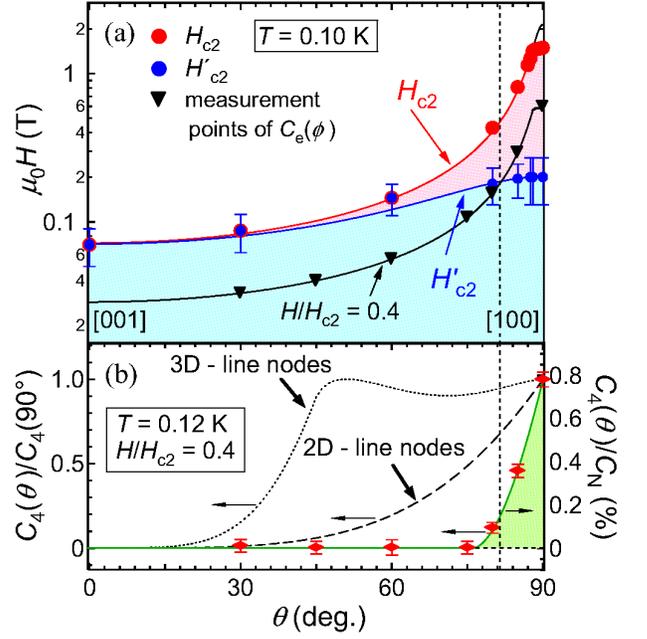}
    \end{center}
\caption{(a) Polar angle $\theta$ dependence of $H_{\rm c2}$ and $H_{\rm c2}^\prime$ at 0.10 K. The red and blue lines represent fits of $H_{\rm c2}$ and $H_{\rm c2}^\prime$ with the Ginzburg-Landau anisotropic effective mass approximation. The black line indicates $H/H_{\rm c2}=0.4$, for which conical azimuthal angle $\phi$ dependence of $C_{\rm e}/T$ is investigated. (b) Polar angle $\theta$ dependence of the normalized 4-fold oscillation amplitude $C_{\rm 4}(\theta)/C_{\rm 4}(90^{\circ})$ and $C_{\rm 4}(\theta)/C_{\rm N}$ in the specific heat. The dotted line and dashed line are calculation results of $C_{\rm 4}(\theta)/C_{\rm 4}(90^{\circ})$.}
\label{fig:HC2ODS}
\end{figure}

	First, to trace the variation of $H^{\prime}_{\rm c2}$ with $\theta$, we investigate the field dependence of $C_{\rm e}/T$ by tilting the polar angle $\theta$. Figure~\ref{fig:THETA}(a) shows the field dependence of $C_{\rm e}/T$ at several polar angles $\theta$ from [100] to [001] directions, fixing the temperature $T = 0.10$ K. For $\bm{H}\parallel [100]$, $C_{\rm e}/T$ increases sharply up to the characteristic field $\mu_{0}H^{\prime}_{\rm c2} = 0.15$ T and then increases more slowly for higher field, which is naturally explained by a presence of two kinds of the gap. By tilting the polar angle $\theta$ from $ab$-plane, these features gradually change in the range of $60^{\circ} < \theta \leq 90^{\circ}$, and finally, the field dependence of $C_{\rm e}/T$ approach $C_{\rm e}/T \propto \sqrt{H}$ in the range of $0^{\circ} \leq \theta \leq 60^{\circ}$. To extract the polar angle dependence of $H^{\prime}_{\rm c2}(\theta)$, we plot in Fig.~\ref{fig:THETA}(b) the field dependence of $C_{\rm e}/T$ against $\sqrt{H/H_{\rm c2}(\theta)}$ at several polar angle $\theta$ from [100] to [001] directions, fixing the temperature $T = 0.10$ K. We directly find the variation of $H^{\prime}_{\rm c2}(\theta)$, marked with arrows in Fig.~\ref{fig:THETA}(b), toward the $H_{\rm c2}(\theta)$ by tilting the polar angle $\theta$ from $ab$-plane in the range of $90^{\circ}$ to $60^{\circ}$: $H^{\prime}_{\rm c2}(\theta) \rightarrow H_{\rm c2}(\theta)$. It is not easy in this way to precisely define $H^{\prime}_{\rm c2}(\theta)$; we will confirm the validity of this definition as described in the next paragraph. In the range of $0^{\circ} \leq \theta \leq 60^{\circ}$, we are able to find good scaling such that $C_{\rm e}/T$ shows universal $\sqrt{H/H_{\rm c2}(\theta)}$ dependence. In this region, we judge $H^{\prime}_{\rm c2}(\theta)$ to be corresponding to $H_{\rm c2}(\theta)$: $H^{\prime}_{\rm c2}(\theta) = H_{\rm c2}(\theta)$.

	Figure~\ref{fig:HC2ODS}(a) shows the polar angle $\theta$ dependence of $H_{\rm c2}$ and $H_{\rm c2}^\prime$, determined from the field dependence of the specific heat at 0.10 K. For orientations not very close to ${\bm H} \parallel$ {\it ab}-plane, the angle dependence of $H_{\rm c2}$ is well fitted with the Ginzburg-Landau anisotropic effective mass approximation.~\cite{Tinkham} Also shown in Fig.~\ref{fig:HC2ODS}(a) is the curve  based on the G-L effective mass model, 
\begin{eqnarray}
H_{{\rm c2}}(\theta) = \frac{H_{{\rm c2} \parallel c}}{\sqrt{\cos^{2}\theta + {{\mathit \Gamma}_a}^{-2}\sin^{2}\theta}}.
\end{eqnarray}
Here $\mathit{\Gamma}_a$ is the square root of the ratio of the effective mass for in-plane motion to that for inter-plane motion (i.e. ${\mathit \Gamma}_a = H_{{\rm c2} \parallel ab}/H_{{\rm c2} \parallel c}$).~\cite{Tinkham} Here we have taken only $H_{{\rm c2} \parallel ab}$ to be the adjustable parameter for the fitting with $H_{{\rm c2} \parallel c}=0.071$ T given. In reality, $H_{\rm c2}$ becomes nearly independent of $\theta$ and strongly undershoots the value obtained from extrapolation of the fitting in the $ 88^{\circ} \le \theta \le 90^{\circ}$ range, due to the limiting of $H_{\rm c2}$ and the occurrence of the second superconducting transition in the $H \parallel$ RuO$_2$ plane.~\cite{Deguchi1} The effective upper critical field of the passive band $H_{\rm c2}^\prime$ can also be fitted without any fitting parameter, using ${\mathit \Gamma}_a^{\prime} = H^{\prime}_{{\rm c2} \parallel ab}/H^{\prime}_{{\rm c2} \parallel c}$, $H^{\prime}_{{\rm c2} \parallel ab}=0.15$ T, and $H^{\prime}_{{\rm c2} \parallel c}=H_{{\rm c2} \parallel c}=0.071$ T. The fitting curve in Fig.~\ref{fig:HC2ODS}(a) indeed confirms that $H^{\prime}_{\rm c2}(\theta)$ defined in Fig.~\ref{fig:THETA}(b) is in accord with the G-L effective mass model.

	The in-plane and inter-plane coherence lengths, $\xi_{ab}$ and $\xi_{c}$, are given in the anisotropic effective mass approximation: $H_{{\rm c2} \parallel c} = \frac{\phi_{0}}{2\pi\xi^{2}_{c}}$, $H_{{\rm c2} \parallel ab} = \frac{\phi_{0}}{2\pi\xi_{ab}\xi_{c}}$. Since the specific heat for ${\bm H}\parallel c$ does not indicate a presence of two kinds of the gap, it is reasonable to assume that $H_{{\rm c2} \parallel c}$ (i.e. $\xi_{ab}$) is common to active and passive bands. In contrast, the plateau in $C_{\rm e}/T$ vs. $H$ at $H_{\rm c2}^\prime$ indicates that $H_{{\rm c2} \parallel ab}$ (i.e. $\xi_{c}$) is different between active and passive bands: $H_{{\rm c2} \parallel ab} = \frac{\phi_{0}}{2\pi\xi_{ab}\xi_{c}}$, $H^{\prime}_{{\rm c2} \parallel ab} = \frac{\phi_{0}}{2\pi\xi_{ab}\xi^{\prime}_{c}}$. The interlayer coherence length is expected to be shorter for the active $\gamma$-band with the dominant quasi-2-dimensional $d_{xy}$ orbital character, thus limiting the main $H_{\rm c2}$. In fact, the quantum oscillations indicate that the $\beta$-band with the $d_{yz}$ and $d_{zx}$ orbital character has the largest interlayer dispersion.~\cite{Bergemann}
\begin{figure}[t]
    \begin{center}
\includegraphics[width=8cm,clip]{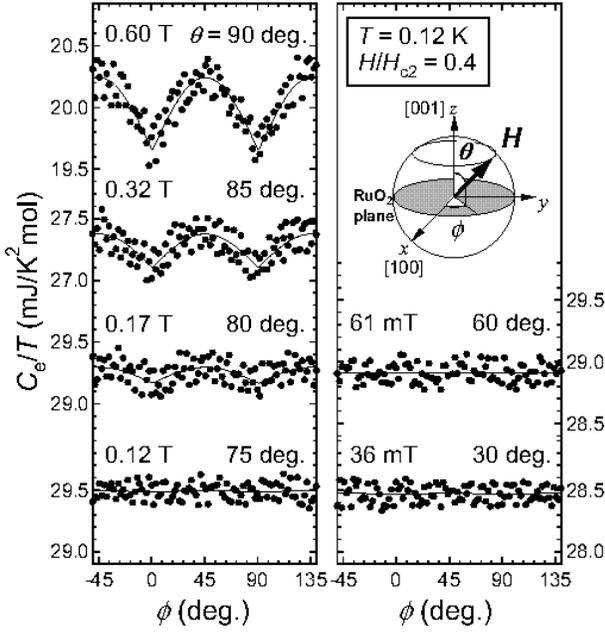}
    \end{center}
\caption{The azimuthal angle $\phi$ dependence of $C_{\rm e}/T$ at several polar angles $\theta$ by rotating ${\bm H}$ conically, fixing the temperature $T = 0.12$ K and the reduced magnetic field $H/H_{c2} = 0.4$. The solid lines are fits with $f_{4}(\phi)$ given in the text.}
\label{fig:OSCth}
\end{figure}

	Next, we examine the polar angle $\theta $ dependence of $C_{\rm 4}(\theta)$ under the condition of $H/H_{\rm c2}(\theta) = 0.4 > H_{\rm th}(\theta)/H_{\rm c2}(\theta)$. Figure~\ref{fig:OSCth} shows $C_{\rm e}/T$ at several polar angle $\theta$, by rotating ${\bm H}$ conically as a function of $\phi$ and keeping $\theta$ constant under the temperature $T = 0.12$ K. The field $H$ at each $\theta$ is adjusted so that the reduced magnetic field is maintained at $H/H_{c2}(\theta) = 0.4$, shown in Fig~\ref{fig:HC2ODS}(a) as triangles. A {\it non-sinusoidal} 4-fold angular variation approximated as $C_{\rm e}(\theta, \phi) = C_{\rm 0}(\theta) + C_{\rm 4}(\theta)f_{\rm 4}(\phi)$ is observed in the range of $80^{\circ} \leq \theta \leq 90^{\circ}$. The oscillation amplitude $C_{4}(\theta)$ decreases steeply with decreasing $\theta$ from $90^{\circ}$ and we cannot detect any angular $\phi$ variation of $C_{\rm e}$ in the range of $\theta < 80^{\circ}$.  The results of the polar angle $\theta$ dependence of $C_{\rm 4}(\theta)/C_{\rm N}$ and $C_{\rm 4}(\theta)/C_{\rm 4}(90^{\circ})$ at $T = 0.12$ K and $H/H_{\rm c2}=0.4$ are summarized in Fig.~\ref{fig:HC2ODS}(b).

	Now, we try to qualitatively estimate the evolution of $C_{\rm 4}(\theta)/C_{\rm 4}(90^{\circ})$ at zero temperature in a single band model. Since we keep the condition of exceeding the threshold field $H_{\rm th}$ at $\theta = 90^{\circ}$: $\frac{H}{H_{\rm c2}(\theta)} > \frac{H_{\rm th}(90^{\circ})}{H_{\rm c2}(90^{\circ})}= 0.1$, the expected result with nodes is qualitatively the same as that with gap minima. Thus we adopt the even-parity SC gap function $\mathit \Delta ({\bm k}) = 2\mathit \Delta_{\rm 0}k_{x}k_{y}$ with vertical line nodes, instead of a realistic odd-parity $\mathit \Delta ({\bm k}) = {\mathit \Delta}_{0}({\sin}ak_x + {\rm i}{\sin}ak_y)$ for which there is no theoretical results available so far. The calculation results of $C_{\rm 4}(\theta)/C_{\rm 4}(90^{\circ})$, based on Refs.~\onlinecite{Vekhter,Won3,Won4}, with 2-dimensional (2D) cylindrical Fermi surface and 3-dimensional (3D) spherical Fermi surface are shown in Fig.~\ref{fig:HC2ODS}(b). To interpret these calculation results, it is essential to remark that the Doppler shift $\delta {\it \omega} = 0$ for $\bm{k} \parallel \bm{H}$. In a 2D cylindrical Fermi surface case, since only the in-plane component of $\bm H$ gives a 4-fold angular variation for $\bm{k} \parallel$ basal plane, $C_{\rm 4}(\theta)/C_{\rm 4}(90^{\circ})$ is gradually suppressed, as $\bm H$ is tilted away from the basal plane and the in-plane component of $\bm H$ is decreased. This behavior agrees with the calculation of point nodes in the basal plane with 3D spherical Fermi surface because the nodal $\bm{k}$ which involves the QP excitations by magnetic fields exists only in the basal plane.~\cite{Izawa4,Thalmeier} In contrast, for vertical line nodes on 3D spherical Fermi surface, the behavior of $C_{\rm 4}(\theta)/C_{\rm 4}(90^{\circ})$ is more complicate and a 4-fold angular variation does not change by a small tilting $\bm H$ from the basal plane. The observed suppression of $C_{\rm 4}(\theta)$ shown in Fig.~\ref{fig:HC2ODS}(b) is much more rapid than any of these single-band expectations.

	Therefore, we need to include the contribution of the SC gap structure in passive bands, in order to explain the polar angle $\theta$ dependence of $C_{\rm 4}(\theta)/C_{\rm 4}(90^{\circ})$.  If the reduction of $C_{\rm 4}$ is caused by compensation of anti-phase anisotropies between active and passive bands, the $\theta$ dependence of $C_{\rm 4}$ is expected to be rapidly suppressed when $H < H^{\prime}_{\rm c2}$ is satisfied, i.e., when both bands become relevant to anisotropic QP excitations. From Fig.~\ref{fig:HC2ODS}(a), the applied field strength, described as the solid line of $H/H_{\rm c2}=0.4$ where $C_{\rm e}(\phi)$ is measured, becomes smaller than $H^{\prime}_{\rm c2}(\theta)$ below $\theta = 81^{\circ}$. This characteristic angle corresponds well with the observed threshold angle of steep suppression of $C_{\rm 4}(\theta)$ in Fig.~\ref{fig:HC2ODS}(b). Thus we conclude that the main origin of steep suppression of $C_{\rm 4}$ at low $T$ and low $H$ is the compensation by gap anisotropy with anti-phase between active and passive bands. Therefore the theoretical models (B)~\cite{Nomura2,Nomura3,Yanase1,Yanase2} is promising to explain the SC gap structure of Sr$_2$RuO$_4$; the SC gap in the passive bands most probably have the lines of gap minima or of zeros in the [110] direction. 

\subsection{Superconducting gap structure of Sr$_2$RuO$_4$}

\begin{table}[t]
\begin{center}
\caption{The classification of typical gap structures in the superconducting state of Sr$_2$RuO$_4$ and consistency with experimental results: Yes (Y) or No (N).}
\label{tab:Full}
\begin{ruledtabular}
\renewcommand{\arraystretch}{1.2}
\begin{tabular}[b]{llccccccc}
\multicolumn{2}{c}{}&\#1&\#2&\#3&\#4&\#5&\#6&\#7\\
\hline
\multicolumn{2}{l}{$\begin{array}[b]{c}{\rm symmetry}\end{array}$}&$p$&$p$&$f$&$f$&$f$&$p+f$&$p$\\
\multicolumn{2}{l}{$\begin{array}[b]{c}{\rm active \; band}\\({\it \gamma})\end{array}$}&\includegraphics[height=0.76cm,clip]{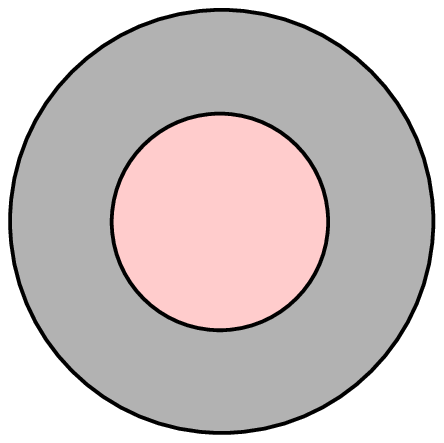}&\includegraphics[height=0.76cm,clip]{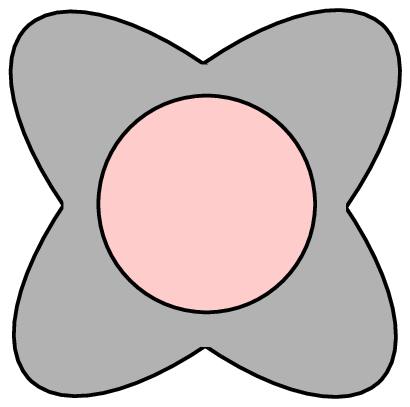}&\includegraphics[height=0.76cm,clip]{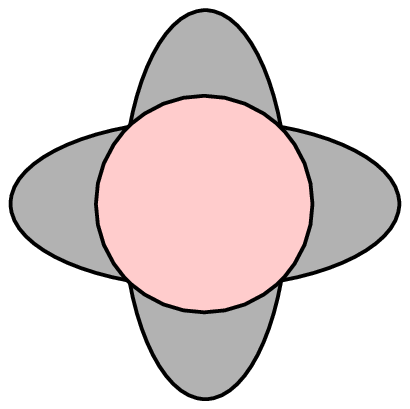}&\includegraphics[height=0.76cm,clip]{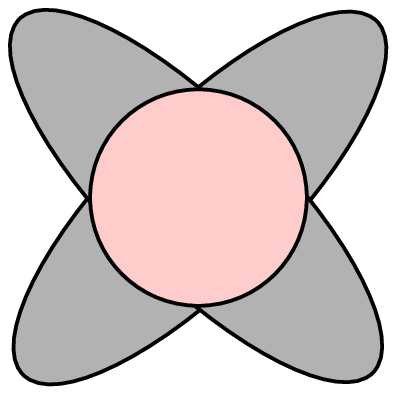}&\includegraphics[height=0.76cm,clip]{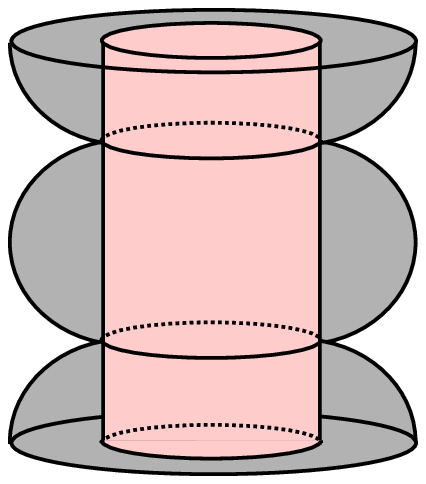}&\includegraphics[height=0.76cm,clip]{MinA.eps}&\includegraphics[height=0.76cm,clip]{MinA.eps}\\
\multicolumn{2}{l}{$\begin{array}[b]{c}{\rm passive \; band}\\({\it \alpha},{\it \beta })\end{array}$}&&&&&&\includegraphics[height=0.74cm,clip]{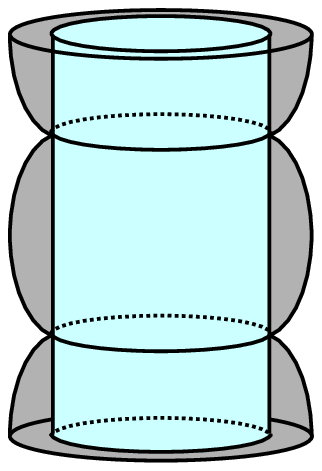}&\includegraphics[height=0.60cm,clip]{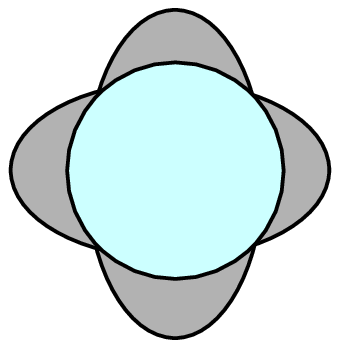}\\
\hline
a.&$C_{\rm e}(T)$&N&N&Y&Y&Y&Y&Y\\
b.&$C_{\rm e}(H)$ $_{H \parallel ab}$&N&N&N&N&N&Y&Y\\
c.&${\it \Delta}C_{\rm e}/{\it \gamma}_{\rm N}T_{\rm c}$&N&N&N&N&N&Y&Y\\
d.&$C_{\rm e}({\it \phi})$&N&Y&N&Y&N&Y&Y\\
e.&$C_{\rm 4}({\it \theta})$&N&N&N&N&N&N&Y\\
f.&Scaling $_{H \parallel c}$&N&(Y)&Y&Y&(Y)&(Y)&Y\\
\hline
\multicolumn{2}{l}{Summary}&N&N&N&N&N&N&Y\\
\multicolumn{2}{l}{Refs.}&\onlinecite{Rice}&\onlinecite{Miyake,Nomura1}&\onlinecite{Kuwabara,Sato,Kuroki,Takimoto,Dahm1,Wu,Eremin}&\onlinecite{Graf}&\onlinecite{Hasegawa1,Hasegawa2,Kubo,Won1}&\onlinecite{Zhitomirsky,Annett1,Annett2,Koikegami}&\onlinecite{Nomura2,Nomura3,Yanase1,Yanase2}\\
\end{tabular}
\end{ruledtabular}
\end{center}
\end{table}
	There have been many theoretical attempts to resolve the controversies on the SC mechanism of the spin-triplet superconductivity in Sr$_2$RuO$_4$ and on the nodal or node-like structure in its SC gap. The SC mechanisms mediated by the anisotropy of the spin fluctuation~\cite{Kuwabara,Sato,Kuroki} or the orbital fluctuation,~\cite{Takimoto} due to the incommensurate AF fluctuation associated with $\alpha$- and $\beta$-bands, were proposed. Consequently, these models lead to the gap structure which corresponds to phenomenologically proposed $f_{x^{2}-y^{2}}$-wave state.~\cite{Dahm1,Wu,Eremin} On the other hand, the spin-triplet superconductivity based on the third order perturbation method for the repulsive Hubbard model for $\gamma$-band, where the effective interaction beyond the spin fluctuation plays important roles, was discussed.~\cite{Nomura1} The anisotropic $p$-wave state derived in this model has essentially the same gap structure as the one derived on the basis of the short-range ferromagnetic interaction.~\cite{Miyake} As other proposals, there are $f$-wave state with horizontal line nodes in the SC gap~\cite{Hasegawa1,Hasegawa2,Kubo,Won1} due to the interlayer pairing interaction and $f_{xy}$-wave state with vertical line nodes.~\cite{Graf}

	Furthermore, several theories,~\cite{Zhitomirsky,Annett1,Annett2,Koikegami,Nomura2,Nomura3,Yanase1,Yanase2} taking the ODS into account,~\cite{Agterberg1} have been proposed. In these models, there are active and passive bands to the superconductivity: the SC instability originates from the active band with a large gap amplitude; pair hopping across active to passive bands induces a small gap in the passive bands. The gap structure with horizontal lines of nodes~\cite{Zhitomirsky,Annett1,Annett2,Koikegami} or strong in-plane anisotropy~\cite{Nomura2,Nomura3,Yanase1,Yanase2} in the passive bands was proposed.
\begin{figure}[t]
    \begin{center}
\includegraphics[width=8cm,clip]{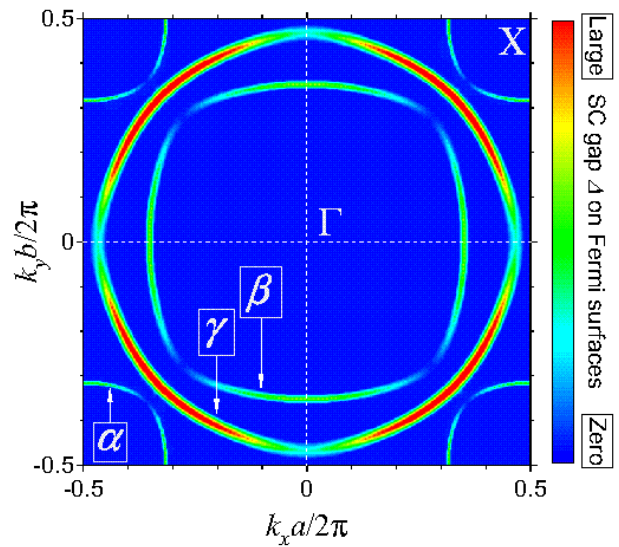}
    \end{center}
\caption{Schematic picture of the SC gap structure of Sr$_2$RuO$_4$ on Fermi surfaces in ${\bm k}$-space. This figure is produced on the basis of the results in Ref.~\onlinecite{Nomura2}.}
\label{fig:GAP}
\end{figure}

	Most of the proposed gap structure can be classified in the five groups (\#1$-$\#5) with the single band models and the two groups (\#6$-$\#7) with the ODS models as summarized in Table~\ref{tab:Full}.  In order to examine the consistency with the experiments, we discussed the items from (a) to (f). The discussions with respect to the items from (a) to (d) are included in our previous report.~\cite{Deguchi2} Now, we try to examine them in the light of the present experimental results and select the appropriate gap structure to fully assign the gap structures in all bands of Sr$_2$RuO$_4$.

 	(a) $C_{\rm e}/T$ in zero field shows $T$-linear dependence down to $T/T_{\rm c} = 1/30$. This result requires the existence of lines of nodes, of zeros, or of deep gap minima in the SC gap in single band models, and in the case of ODS, the SC gap of the passive bands should have nodes, zeros, or gap minima. Thus \#1 with an isotropic gap and \#2 with shallow gap minima in single band models are inappropriate. (b) At low temperature, $C_{\rm e}/T$ for ${\bm H} \parallel ab$-plane reveals the typical field dependence with two kinds of the gap and supports the ODS models \#6 and \#7 with the active band $\gamma$ and the passive bands $\alpha, \beta$. Next we address (c) the specific heat jump ${\mathit \Delta}C_{\rm e}/\gamma_{\rm N}T_{\rm c}$ at $T_{\rm c}$ in zero field. The theoretical calculations of ${\mathit \Delta}C_{\rm e}/\gamma_{\rm N}T_{\rm c}$ with the weak coupling gives 1.43 for \#1, 1.22 $-$ 1.07 (for ${\mathit \Delta}_{\rm min}/{\mathit \Delta}_{\rm max} = 1/2 - 1/4$) for \#2, and 0.95 for \#3$-$\#5 in single bands models, respectively, and thus overestimate the experimental result ${\mathit \Delta}C_{\rm e}/\gamma_{\rm N}T_{\rm c} = 0.73$. On the other hand, the ODS models of \#6 and \#7, which have the gap structure of \#2 for the active band $\gamma$, give good agreement: ${\mathit \Delta}C_{\rm e}/\gamma_{\rm N}T_{\rm c} = (1.22 - 1.07) \times 0.57 = 0.70 - 0.61$, mainly originating from the active band $\gamma$ with the {\it gap minimum} (${\mathit \Delta}_{\rm min}/{\mathit \Delta}_{\rm max} = 1/2 - 1/4$). We note that a very deep gap minima in the active band will lead to smaller ${\mathit \Delta}C_{\rm e}/\gamma_{\rm N}T_{\rm c}$ than the observation.

	(d) For the field range $0.15$ T $< \mu_{0}H < 1.2$ T and ${\bm H} \parallel ab$-plane, where the QPs in the active band $\gamma$ are the dominant source of in-plane anisotropy in $C_{\rm e}(\phi)$, we deduce the existence of a node or of a gap minimum along the [100] direction, because $C_{\rm e}$ takes a minimum with the 4-fold oscillation. This result places severe constraints on the gap structure and requires the SC gap with the 4-fold anisotropy of \#2 or \#4 in active band, but not of \#3. (e) Polar angle $\theta$ dependence of the in-plane 4-fold oscillation $C_{\rm 4}(\theta)$ gives an answer to the disappearance of the in-plane 4-fold oscillation of $C_{\rm e}(\phi)$ below 0.15 T. The issue was whether it originates solely from the gap minima of the active band $\gamma$ or mainly from the compensation for the field dependent QP excitations from the passive bands $\alpha, \beta$. Steep suppression of $C_{\rm 4}(\theta)$ by tilting the polar angle $\theta$ from the \RO\ plane is explained only by the compensation from the passive bands $\alpha, \beta$ with the ODS model \#7. Finally, (f) the scaling of $C_{\rm e}(T, H)$ for ${\bm H} \parallel c$-axis confirms the existence of {\it lines} of nodes, of zeros, or of gap minima along the $k_{z}$ direction in the SC gap of both active and passive bands.

	To summarize consistencies with experiments on the items (a)$-$(f), we conclude that the ODS model \#7 is suitable for description of the spin-triplet superconductivity of Sr$_2$RuO$_4$. Figure~\ref{fig:GAP} depicts the SC gap structure of Sr$_2$RuO$_4$ on Fermi surfaces in ${\bm k}$-space, based on Ref.~\onlinecite{Nomura2}. If the essential order parameter is with the $\hat {\bm{z}}(k_x + {\rm i}k_y)$ symmetry in the active band, the origins of the SC gap structure of \#7 for each Fermi surface is explained qualitatively as follows. The order parameter has nodes at $(\pm \pi, 0)$ and $(0, \pm \pi)$ in the $k$-space due to the periodicity and odd parity. For the active band $\gamma$, whose Fermi surface passes in the vicinity of these nodal points, the SC gap consequently has the lines of gap minima in the [$\pm$100] and [0$\pm$10] directions; thus the order parameter is described as $\bm{d}(\bm{k})=\hat {\bm{z}}{\mathit \Delta}_{0}({\sin}ak_x + {\rm i}{\sin}ak_y)$.~\cite{Nomura1,Miyake} In the passive bands $\alpha$ and $\beta$, the superconductivity with the same SC symmetry is coherently induced through the inter-band Coulomb interaction. However, the strong incommensurate AF fluctuations by the nesting properties of the Fermi surfaces of $\alpha$- and $\beta$-bands, with the nesting vector in the diagonal directions,~\cite{Sidis,Braden1,Braden2,Mazin2} give unfavorable contributions to support this superconductivity with the $\hat {\bm{z}}(k_x + {\rm i}k_y)$ symmetry.~\cite{Nomura2,Nomura3,Yanase1,Yanase2} Thus the SC gap in the passive bands have the lines of gap minima or zeros to the [$\pm$1$\pm$10] and [$\pm$1$\mp$10] directions to avoid the contributions of the incommensurate AF fluctuations, which would favor $d$-wave. Therefore the experimental results are consistent with the models of \#7, in which the order parameter with the $\hat {\bm{z}}(k_x + {\rm i}k_y)$ symmetry in the active band originates from the pairing interaction which favors {\it the odd parity orbital state}, not from the ferromagnetic or antiferromagnetic spin fluctuations.

\section{Conclusion}
In conclusion, we have obtained strong evidence for full determination of the superconducting gap structure of the spin-triplet superconductor Sr$_2$RuO$_4$ in all bands for the first time, based on the field-orientation dependence of specific heat at low temperature and the scaling of electronic specific heat in magnetic fields. We confirmed the multi-band superconductivity with the active band $\gamma$ and the existence of a modulated superconducting gap with the line of a minimum along the [100] direction in the active band. This gap structure in the active band is well described with $p$-wave order parameter $\bm{d}(\bm{k})=\hat {\bm{z}}{\mathit \Delta}_{0}({\sin}ak_x + {\rm i}{\sin}ak_y)$. Furthermore, in the passive bands $\alpha$ and $\beta$, the existence of the line of a minimum or of a zero in the induced superconducting gap to the [110] direction is experimentally indicated. This result suggests the incommensurate antiferromagnetic fluctuations at $\alpha$- and $\beta$-bands gives unfavorable contributions to this superconductivity. Among the superconducting mechanisms proposed so far for Sr$_2$RuO$_4$, the ones based on the vertex correction of Coulomb repulsion, other than spin fluctuations, best reproduce the experimental observations.

\begin{acknowledgements}
We thank N. Kikugawa, T. Nomura, H. Ikeda, Kosaku Yamada, K. Machida, T. Ohmi, M. Sigrist, K. Ishida, H. Yaguchi, A.P. Mackenzie, S.A. Grigera, K. Maki, and T. Ishiguro for stimulating discussions and comments. We are particularly grateful to T. Nomura and Kosaku Yamada for kindly permitting us to use the numerical results for the gap structure. This work was in part supported by the Grants-in-Aid for Scientific Research from JSPS and MEXT of Japan, by a CREST grant from JST, and by the 21COE program ``Center for Diversity and Universality in Physics'' from MEXT of Japan. K. D. has been supported by JSPS Research Fellowship for Young Scientists.

\end{acknowledgements}

\end{document}